[Author Name(s)]

Yaolong Zhang, Jun Jiang, and Bin Jiang

[Author Affiliations(s)]

*Hefei National Laboratory for Physical Science at the Microscale, Department of Chemical Physics, University of Science and Technology of China, Hefei, Anhui 230026, China*

**[Chapter Starts Here]**

**Abstract:**

Machine learning of scalar molecular properties such as potential energy has enabled widespread applications. However, there are relatively few machine learning models targeting directional properties, including permanent and transition dipole (multipole) moments, as well as polarizability. These properties are essential to determine intermolecular forces and molecular spectra. In this chapter, we review machine learning models for these tensorial properties, with special focus on how to encode the rotational equivariance into these models by taking a similar form as the physical definition of these properties. You will then learn how to use an embedded atom neural network model to train dipole moments and polarizabilities of a representative molecule. The methodology discussed in this chapter can be extended to learn similar or higher-rank tensorial properties, such as magnetic dipole moments, non-adiabatic coupling vectors, and hyperpolarizabilities.

[H1 Introduction]

Modern quantum chemistry theories offer various ways of solving electronic Schrödinger equations. These solutions yield not only the total potential energy, but also the corresponding wavefunction and electron density, from which a diversity of physicochemical properties of molecules and materials can be derived. With the aid of machine learning (ML), accurate and efficient high-dimensional representations of potential energy[1] and electron density[2] have emerged in recent years, which have accelerated first-principles molecular simulations in a wide range of systems by several orders of magnitude[3-4]. Such advances have been reviewed in several chapters of this book. While in this chapter, we focus on ML of (permanent or transition) dipole moments and related properties, such as multipoles and polarizabilities, all can be obtained from quantum chemical calculations. These properties are essentially responsible for the interaction between light and matter at different levels[5], which can be used for computing infrared spectra (dipole moment), Raman spectra (polarizability), and electronic spectra (transition dipole moment, as will be seen in the next chapter), etc. On the other hand, they are often useful for evaluating the long-range electrostatic, induction, and dispersion contributions to intermolecular forces[6]. As a result, learning these properties is of great importance both in spectroscopic simulations and in the construction of force fields with inclusion of long-range interactions.

It is important to note that the potential energy, dipole/multipole moment, and polarizability tensor have different behaviors under symmetry operations. As a scalar property, the potential energy is invariant with respect to overall translation, rotation, and permutation of identical atoms of the system. In comparison, the permanent dipole moment is a measure of charge separation in a system, which is a vector classically expressed by the separated charge ($q$) multiplied by the vector between the positive and negative charges ($\vec{r}$), i.e. $\vec{\mu}=q\vec{r}$. In a more general description, electric multipole moments are correlated with the multipole expansion of charge density, where dipole moment ($\vec{\mu}$) is the first-order term and there are second-order (quadrupole, $\Theta$), third-order (octupole, $\Omega$) moments, and so on and so forth. For example, $\vec{\mu}$ and $\Theta$ are first- and second-rank tensors with the following components[7],

$$\mu_\alpha = \sum_{i=1} q_i r_{i\alpha}, \tag{1}$$

$$\Theta_{\alpha\beta} = \frac{1}{2}\sum_i q_i(3r_{i\alpha}r_{i\beta} - |\mathbf{r}_i|^2 \delta_{\alpha\beta}), \tag{2}$$

where $q_i$ is the $i^{th}$ point charge, $\mathbf{r}_i$ is its position relative to a certain origin (typically the center of mass or the central atom in the atomistic ML framework) in the molecule, with the Greek subscripts denoting $x$, $y$, $z$ components, $\delta_{\alpha\beta}=1$ if $\alpha\neq\beta$, otherwise $\delta_{\alpha\beta}=0$. These multiple moments determine the weak interaction of a molecule with an external field. In addition to these ground state response properties, it should be noted that the transition dipole moment associates with the transition between two different electronic states, and is computed from the integral over electronic coordinates of the dipole moment operator with respective wavefunctions,

$$\vec{\mu}_{if}^T = \langle \psi_f | \hat{\mu} | \psi_i \rangle_e, \tag{3}$$

where $\psi_i$ and $\psi_f$ correspond to the initial and final state wavefunctions, respectively. Since the transition dipole moment reflects the change of charge distribution upon electronic transition, its vectorial direction depends on the transition type (e.g. parallel or perpendicular to the molecular plane) and its sign depends on the relative phase of the initial and final state wavefunctions, resulting in additional complexity for ML.

On the other hand, the polarizability ($\boldsymbol{\alpha}$) is a second-rank tensor as a result of the second-order response of potential energy ($V$) to an electric field ($\vec{\mathbf{E}}$), reflecting the ability of the distortion of the charge distribution by the field,

$$\boldsymbol{\alpha} = \frac{\partial^2 V}{\partial \vec{\mathbf{E}}^2}. \qquad (4)$$

While the polarizability can depend on the field strength, we are often interested in the static polarizability at $\vec{\mathbf{E}} = 0$. Apparently, these tensorial properties contain multiple coordinate-dependent components that are translationally and permutationally invariant, yet equivariant with respect to rotation, which means that rotating the molecule will lead to an equivalent rotation to the whole tensor. Next, we will review various approaches to deal with these issues and emphasize the essence of mimicking the physical definition of each property in developing relevant ML models.

# Methods

## Learning permanent and transition dipole moments

Historically, dipole moment has long been interpolated/fitted in a similar way as constructing potential energy surfaces (PESs). For example, global permanent and transition dipole moment surfaces of small molecules have been often interpolated in internal coordinates suitable for quantum dynamics calculations of infrared and electronic spectra[8-16]. While early applications relied on the spline interpolation[8-11], neural networks (NNs) have been more frequently applied in recent years[12-16]. This strategy requires the molecule to be placed in a fixed body-fixed frame when performing ab initio calculations and fits individual dipole moment components independently, which is very system-specific and does not guarantee symmetry rigorously. Likewise, it remains a common way for bigger systems to manually align all configurations to a local reference frame so as to learn the one-to-one correspondence between molecular structures and individual components of a tensorial property[17-24]. Specifically, Popelier and coworkers proposed to replace the empirical functions with machine learned atomic multipole moments for specific molecules in terms of internal coordinates[17,22], with the aim to better model the electrostatic long-range interactions in polarizable force fields[23]. This idea was later extended to learn a transferable model of atomic multipole moments for a wide range of molecules with the Coulomb-matrix descriptor[18]. Marquetand and coworkers [21,24] and Jiang and coworkers[20,25-26] instead focused on learning permanent and transition dipole moments with various structural descriptors and applying machine-learned properties to boost the simulations of ultraviolet and infrared spectra. Despite these successful applications, this strategy is less well-defined in heavily-distorted or dissociable systems and could cause discontinuity at the boundary of symmetry operation.

Taking the rotational equivariance of a tensorial property fully into account is necessary to obtain a continuous and accurate representation. The first properly symmetry-adapted representation of permanent dipole moment can be perhaps traced back to the pioneering work of Bowman and coworkers[27] using the so-called permutationally invariant polynomials (PIPs)[28-29] approach. These PIPs are symmetrized products of functions of relevant interatomic distances according to the invariant theory[28], which are invariant with respect to permutation of any identical atoms and thus suitable for PES construction. To fit dipole moments, the authors expand the dipole moment as the sum of atomic contributions with the assumption that electronic contributions in Eq. (1) have been integrated out and concentrated at individual atomic sites. In such so-called "atomic charge" or "partial charge" representation, one only needs to fit the scalar atomic charge and the vectorial property is guaranteed by the atomic position vector. However, it is important to realize that the charges of two identical atoms should be exchanged under the permutation operation, while the charge of any unique atom should remain unchanged. In practice, therefore, effective atomic charges of unique atoms will be expressed by invariant polynomials with full permutation symmetry, while those of identical atoms have to take a lower symmetry that distinguish relevant atoms making the resultant PIPs interchangeable with respect to permutation[30]. This adaption successfully encodes the correct equivariant property and permutation symmetry of the dipole moment and has been widely applied in learning permanent dipoles of small molecules[31-33]. The same trick was adopted in the PIP-NN approach developed by Guo and coworkers[34-35] to develop several symmetry-preserving permanent dipole surfaces[36-37]. This approach was further extended by Medders and Paesani to derive formally exact many-body expansions of the dipole moment and polarizability for liquid water[38].

The drawback of using a molecule-wise descriptor like PIPs is that its complexity increases dramatically with the size of the molecule. Indeed, on the basis of the atomic charge model, it is more natural to develop an atom-wise ML representation for the dipole moment. The first successful atomistic ML model was proposed by Behler and Parrinello in 2007 for representing high-dimensional PESs[39]. The core idea later adopted in many ML potentials (see the first three Chapters of Part 2) is to decompose the total energy into atomic energies. For a system with $N$ atoms,

$$E = \sum_{i=1}^{N} E_i, \qquad (5)$$

each of which depends on the respective atom-centered local environment only. Each atomic energy is a scalar output of an atomic NNs whose parameters are element-dependent, making the total energy permutationally invariant. While this Behler and Parrinello type of NN (BPNN) model was proposed to learn total energy since atomic energies are not meaningful outcomes of quantum chemistry, it was soon extended to predict atomic environment-dependent charges[40] computed by the Hirshfeld charge partitioning scheme.[41] Note that atomic charges of identical atoms, like atomic energies, are naturally interchangeable in this BPNN model, thus avoiding the more intricate symmetrization in PIPs. This idea was later taken by Rai and Bakken[42], who instead used random forest to fit atomic charges to reproduce the quantum chemical electrostatic potential and then retrieved dipole moments using Eq. (1).

These studies have laid the foundation for later ML models of tensorial properties, with the wide acceptance of the atomistic concept in the community. To avoid the arbitrariness in any charge partitioning scheme, Marquetand and coworkers[43] first proposed to directly train BPNNs to reproduce dipole moment (like the total energy), in which the environment-dependent atomic charges are inferred as latent variables (like atomic energies), by minimizing the following cost function,

$$S(\mathbf{w}) = \sum_{m=1}^{N_{data}} [(Q_m^{NN} - Q_m^{Ref})^2 + |\vec{\mu}_m^{NN} - \vec{\mu}_m^{Ref}|^2 / 3] / N_{data}. \qquad (6)$$

Here $N_{data}$ is the number of data points in the training set, $NN$ and $Ref$ in superscripts refer to the values of NN output and the target. Note that $Q$ is the total charge of the system (*e.g.* $Q=0$ for neutral species) and $Q^{NN} = \sum_i^N q_i^{NN}$, which is actually not a necessity in Eq. (6) if the dipole moment is the only quantity of interest. In such a scheme, the only difference between training dipole moment and potential energy is that now the output of atomic NNs is first multiplied with the atomic position vector and then summed over to generate the total dipole moment. This strategy can be easily extended to include higher order multipole moments[43].

This atomic charge model has been adapted with other atomistic ML models to learn permanent dipole moments[44-47]. This is made by default in some NN packages, such as TensorMol[48] and PhysNet[49], so as to conveniently take long-range electrostatic interactions into account. It has also been recently extended to learn transition dipole moments for a number of excited states of several molecules, with which the UV spectra of similar molecules yet not included in the training set were reasonably reproduced[50]. However, being an excited-state property, learning transition dipole moment is more involved than learning the permanent counterpart (see next Chapter *Learning excited-state properties for more details*). First, since the phase of quantum wavefunction is always arbitrary, so is the sign of transition dipole moment, a proper phase correction needs to be done before using any kind of ML model. While this is not a trivial task, Marquetand et al. have proposed a pragmatic procedure for phase correction by computing wave function overlaps between new geometries and a reference one (*e.g.*, equilibrium geometry), with which the complete training set can be phase corrected[21]. The same group further developed an automatic phase correction program by enabling the sign flip during training and chose the proper one with lower fitting error[51]. Second, it is less recognized that the direction of transition dipole moment also depends on the molecular orbitals involved in the transition. For example, the transition dipole moment vector of a planar molecule will be in plane for parallel transition, while out of plane for perpendicular transition. However, since all atoms lie in the plane, none of atomic dipoles defined in Eq. (1) will have out-of-plane components. As a consequence, the simple atomic charge model would completely fail to predict a perpendicular transition dipole moment. It should be noted that this influence is not limited to pure planar geometries. Indeed, such deficiency will certainly introduce a distortion in the feature space, which hinders the ability to perform the overall regression of transition dipole moments. To solve the problem, we introduce two analogous vectors to the permanent dipole[45], namely,

$$\vec{\mu}_T^j = \sum_{i=1}^{N} q_i^j \vec{r}_i \quad (j=1, 2), \tag{7}$$

where $q_i^j$ ($j$=1, 2) can be obtained by two different outputs of the same atomic NN and the directions of $\vec{\mu}_T^1$ and $\vec{\mu}_T^2$ are automatically determined by NNs. As long as $q_i^1$ and $q_i^2$ are not accidentally identical, $\vec{\mu}_T^1$ and $\vec{\mu}_T^2$ will define a specific plane (*e.g.*, the molecular plane for a planar geometry) and their cross product will give rise to a third vector perpendicular to this plane,

$$\vec{\mu}_T^3 = \sum_{i=1}^{N} q_i^3 (\vec{\mu}_T^1 \times \vec{\mu}_T^2), \tag{8}$$

where $q_i^3$ is given by another output of the same atomic NN that determines the magnitude of $\vec{\mu}_T^3$. These three fundamental vectors are then linearly combined, namely $\vec{\mu}_T^{NN} = \vec{\mu}_T^1 + \vec{\mu}_T^2 + \vec{\mu}_T^3$, to reproduce the direction of transition dipole moment with the correct rotational equivariance. Here we keep using the symbol $q_i^j$ for consistency, which however has no physical meaning. This approach has successfully reproduced the both perpendicular and parallel transition dipole moments of N-methylacetamide[45].

[H2 Learning polarizabilities]

Compared to the large amount of work for learning dipole moment as discussed above, ML models that fully account for the symmetry of higher-rank tensorial properties are scarce. Based on Gaussian process regression, Ceriotti and coworkers have first made use of a tensorial kernel function in spherical coordinates to describe the structural similarity between molecular configurations and meanwhile account for the rotational equivariance of the tensorial properties up to arbitrary order[52-54]. This so-called symmetry-adapted Gaussian process regression (SA-GPR) model for polarizability tensor has been applied to simulate the vibrational Raman spectra of Paracetamol in various crystal forms[55]. By effectively learning an atom-centered model of the polarizability, the SA-GPR scheme has been used to predict the polarizability in chemical space that are transferable between different molecules.[53] Taking tensorial kernel functions is sufficient to preserve the desired symmetry of the kernel-based tensorial representation, because its output is essentially a liner product of tensorial kernels. However, using the same strategy to construct the rotationally-equivariant input would not work for conventional NNs, because feeding such an input into an arbitrary nonlinear function like NNs will generally lead to predictions that do not satisfy the desired symmetry properties.[54]

Following the definition in Eq. (4), we recently modified the atomic charge NN framework[45] to learn the polarizability tensor. To this end, we design an auxiliary matrix (**D**) given by the partial derivatives of atomic charge with respect to atomic coordinates (analogous to the atomic force matrix, in $3 \times N$ dimension),

$$\mathbf{D}^j = \sum_{i=1}^{N} \frac{\partial q_i}{\partial \vec{\mathbf{r}}_j} \quad (j = 1, 2, \ldots, N). \tag{9}$$

Here, $\mathbf{D}^j$ is the $j$th column of **D** matrix. Multiplying the **D** matrix by its transpose gives us a symmetric $3 \times 3$ matrix that has the same transformation as **α**,

$$\boldsymbol{\alpha}^{\mathrm{NN1}} = \mathbf{D}(\mathbf{D})^T. \tag{10}$$

However, $\boldsymbol{\alpha}^{\mathrm{NN1}}$ is a semidefinite matrix by construction, while molecular polarizability itself is not necessarily semidefinite. We thus create another symmetric matrix $\boldsymbol{\alpha}^{\mathrm{NN2}}$ in the following way,

$$\boldsymbol{\alpha}^{\mathrm{NN2}} = \mathbf{r}(\mathbf{D})^T + \mathbf{D}\mathbf{r}^T, \tag{11}$$

where **r** is the 3×N Cartesian coordinate matrix. Furthermore, it is important to note that both $\boldsymbol{\alpha}^{NN1}$ and $\boldsymbol{\alpha}^{NN2}$ become a rank-deficient matrix when the molecular geometry is planar, while the molecular polarizability tensor is not. The simplest way to correct this is to incorporate a scalar matrix $\boldsymbol{\alpha}^{NN3}$, whose diagonal elements are identical which can be thus represented by a scalar EANN (just like the potential). Combining these three terms yields the full representation of the NN-based polarizability tensor,

$$\boldsymbol{\alpha}^{NN} = \boldsymbol{\alpha}^{NN1} + \boldsymbol{\alpha}^{NN2} + \boldsymbol{\alpha}^{NN3}, \quad (12)$$

that fulfills the symmetry of **α**. This approach is in principle applicable to any atomistic ML model with symmetry-invariant descriptors.

A similar approach was independently developed by Sommers *et al.*[56] based on the Deep potential MD (DPMD) framework. Instead of using specific symmetry functions, DPMD defines $N_c$ element-wise embedding NNs ($N_c$ is the number of neighbor atoms), each of which converts the internuclear distance between a neighbor atom and the central atom to $M$ outputs. This leads to a matrix **E** ($M \times N_c$) multiplied with a generalized Cartesian coordinate matrix **r** ($N_c \times 4$) (or a regular Cartesian coordinate matrix **r** ($N_c \times 3$)) to obtain a rotationally equivariant matrix **T** ($M \times 4$) (or $M \times 3$), namely,

$$\mathbf{T} = \mathbf{Er}. \quad (13)$$

The product of **T** and its transpose $\mathbf{T}(\mathbf{T})^T$ generates $M \times M$ invariant atomic features that can be applied as the input of subsequent atomic NNs for learning atomic scalar properties. To learn the polarizability tensor and preserve its rotational equivariance, the atomic polarizability is represented by a diagonal matrix **F** ($M \times M$) inserted in between **T** ($M \times 3$) and its transpose $(\mathbf{T})^T$ ($3 \times M$),

$$\alpha_i^{\mathrm{NN}} = (\mathbf{T})^T \mathbf{F} \mathbf{T}. \tag{14}$$

Diagonal elements of **F** correspond to $M$ scalar outputs of regular DPMD. More details on the DPMD method may be found in the original publication[56] and Chapter 13: *Dynamics with neural network potentials*. In practice, this expression has more or less a combinative effect of Eqs. (10) and (11). Sommers *et al.*[56] have applied this model to liquid water and simulated the Raman spectra by machine-learned polarizabilities. However, this approach will in principle suffer from the aforementioned rank deficiency issue for a planar molecule, since each row of **T** resembles a vector in the same plane in this case.

[H2 Other relevant approaches]

Significant efforts have also been made in developing equivariant message-passing type of neural networks (MPNNs), which can preserve directional information by making the neurons not only scalars but also vectors (tensors) that are internally equivariant to rotation[57-58]. Very recently, equivariant MPNNs have been adapted to predict dipole moments and polarizability tensors enabling accurate infrared and Raman spectra from MD simulations[59], as well as multipole moments to describe electrostatic interaction[44]. Interestingly, a similar expression as Eq. (12) was used by Schütt et al. in their MPNN model for learning polarizability, who have rationalized that the first two terms are related with the anisotropic components of the polarizability tensor, while the last term corresponds to the isotropic component[59].

We note that there are some alternative approaches for learning tensorial properties differing from those discussed above. Within kernel-based regression, for example, Christensen *et al.* applied the response operator to kernel functions to mimic the computation of the response properties of potential energy to an electric field, which explicitly incorporates electric field gradients into the training[60]. This is perhaps the only ML model that yields field-dependent quantities. On the other hand, one may directly learn the equivariant ground state electronic wavefunctions (see Chapter *Learning wavefunction*), *e.g.* using the SchOrb model[61-62], from which the dipole moment, polarizability and related quantities can be all obtained.

[H2 Learning tensorial properties with embedded atom neural networks]

Before we move on to case studies, we describe the details of learning tensorial properties with our recently developed embedded atom neural network (EANN) model[45,63]. This model is inspired from the empirical embedded atom method (EAM) force field which was originally applied to metals[64], also based on the atomic energy decomposition in Eq. (5). In particular, the EANN model uses embedded electron density features ($\boldsymbol{\rho}_i$) at the central atom position as descriptors for the local environment. Each atomic NNs serves as the mapping function from embedded density to the atomic energy. For simplicity, the embedded density feature can be evaluated by the square of the linear combination of neighbor atomic orbitals,

$$\rho_i = \sum_{l_x,l_y,l_z}^{l_x+l_y+l_z=L} \frac{L!}{l_x!l_y!l_z!} \left[ \sum_{j \neq i}^{N_c} c_j \varphi(\hat{\mathbf{r}}_{ij}) f_c(r_{ij}) \right]^2, \quad (15)$$

where $\hat{\mathbf{r}}_{ij} = \hat{\mathbf{r}}_i - \hat{\mathbf{r}}_j$, with $\hat{\mathbf{r}}_i = (x_i, y_i, z_i)$ and $\hat{\mathbf{r}}_j = (x_j, y_j, z_j)$ being the Cartesian coordinate vectors of the central atom *i* and a neighbor atom *j*, $r_{ij} = |\hat{\mathbf{r}}_{ij}|$ is the distance between them, $\varphi(\hat{\mathbf{r}}_{ij})$ is the Gaussian-type orbital centered at atom $j$ parameterized by its center ($r_s$), width ($\alpha$), and angular momenta ($L= l_x+l_y+l_z$),

$$\varphi(\hat{\mathbf{r}}_{ij}) = (x_i - x_j)^{l_x}(y_i - y_j)^{l_y}(z_i - z_j)^{l_z} \exp\left[-\alpha\left(r_{ij} - r_s\right)^2\right], \qquad (16)$$

$f_c(r_{ij})$ is a cutoff function continuously damping the invariant to zero at the cutoff radius ($r_c$), and $N_c$ is the number of atoms within $r_c$. Clearly, $\rho_i$ corresponds to the embedded density contribution from a given type of atomic orbital and expresses two- ($L=0$) and three-body ($L>0$) interactions in a uniform way. Indeed, Eq. (15) allows the evaluation of atom-centered three-body terms at a cost of atom-centered two-body ones, resulting in a linear scaling with respect to the number of neighbors. This EANN model has enabled accurate and efficient learning of potential energies and forces[65-70], as well as Hessians[71]. It has been found to be more efficient than many other descriptor-based ML models[67] and can be further improved to have a message-passing feature[72].

# Case studies

In this section, you will see how to use machine learning for learning permanent and transition dipole moments, as well as polarizability tensors. We use the N-methylacetamide (NMA) molecule as a toy system, which is a representative residual in protein backbones[20]. Permanent dipole moments and polarizabilities were calculated by density functional theory (DFT). The perpendicular nπ* and parallel ππ* transition dipole moments, relative to the molecular surface as illustrated in Figure 1, were calculated by time-dependent density functional theory (TDDFT)[73]. All these calculations were performed using the Gaussian 09 package at the B3LYP/cc-pVDZ level[74]. We extracted 10000 NMA configurations from 1000 different protein backbones at room temperature from the Protein Data Bank (PDB)[75], among which 80% randomly selected structures were used for training and the rest were used for testing. Phase corrections have been carefully done by evaluating the wavefunction overlaps between these structures and a reference configuration. These data can be found at https://github.com/zylustc/Learning-Dipole-Moments-and-Polarizabilities.

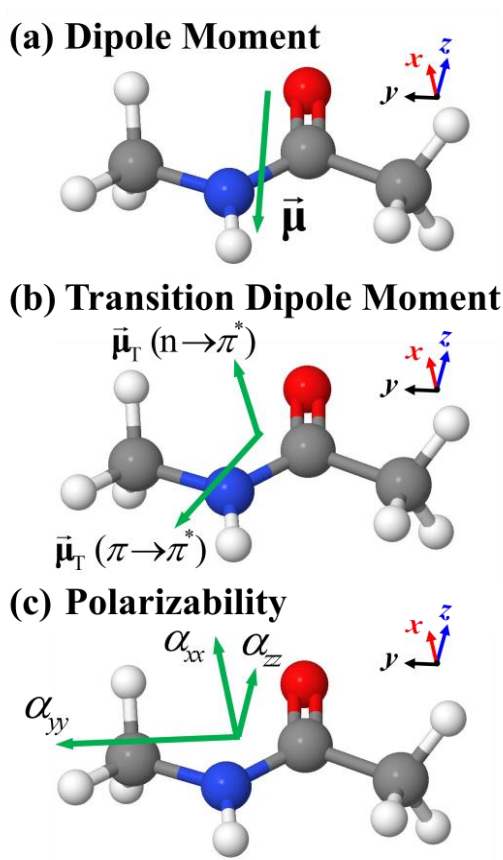

*** Insert Figure 1 ***

Caption: Schematic diagrams of the equilibrium geometry of NMA and its corresponding (a) permanent dipole moment, (b) transition dipole moments and (c) polarizability. The arrows in (a) and (b) represent the directions of the dipole moments and those in (c) are the isotropic polarization ($α_{xx}$, $α_{yy}$, $α_{zz}$) in each direction.

You can learn these properties with the EANN code based on Pytorch[76], which can be found at https://github.com/zhangylch/EANN/. All well-trained EANN dipole and polarizability models in this work are freely available at "https://github.com/zylustc/Learning-Dipole-Moments-and-Polarizabilities". A script named "test.py" showing how to predict these tensorial properties is given in the "readme" of the github repository. This script will load a saved model and pass the required information to obtain the desired tensorial properties, for example the model in the 'potentials/dipole moment' folder for predicting permanent dipole moments. In our tests, thirty-three embedded density features with $L$ up to 2 were used to depict the local environment and NN structures consisting of three hidden layers (64×32×16) were employed to map the embedded density to the atomic charge, respectively. These settings are identical for permanent and transition dipole moments, as well as polarizability.

Figure 2 displays the correlation diagrams between the DFT determined properties and the EANN predictions of the NMA molecule. For permanent dipole vectors and polarizability tensors, the test root mean square of errors (RMSEs) are merely 0.004 a.u. and 0.020 a.u., respectively, and the correlation coefficients ($r$) are essentially unity. It should be noted that the elements in these tensorial properties scatter in rather different numerical ranges, all of which are well-reproduced by our symmetry-preserving tensorial EANN models. Figs. 2c and 2d further show the performance of the tensorial EANN model in predicting transition dipole moments. In particular, our model properly describes the perpendicular n-π* transition of NMA with a very low RMSE (0.002 a.u.) which consists mainly of the transition of the near planar peptide bond (HN-C=O). Note that ML based on Eq. (1) only would fail in this case, leading to considerably larger errors.[45] This is because the electronic transition in the NMA molecule primarily involves the C=O-N-H group, which is basically a planar geometry. Likewise, the prediction of the EANN model for π-π* is also equally accurate.

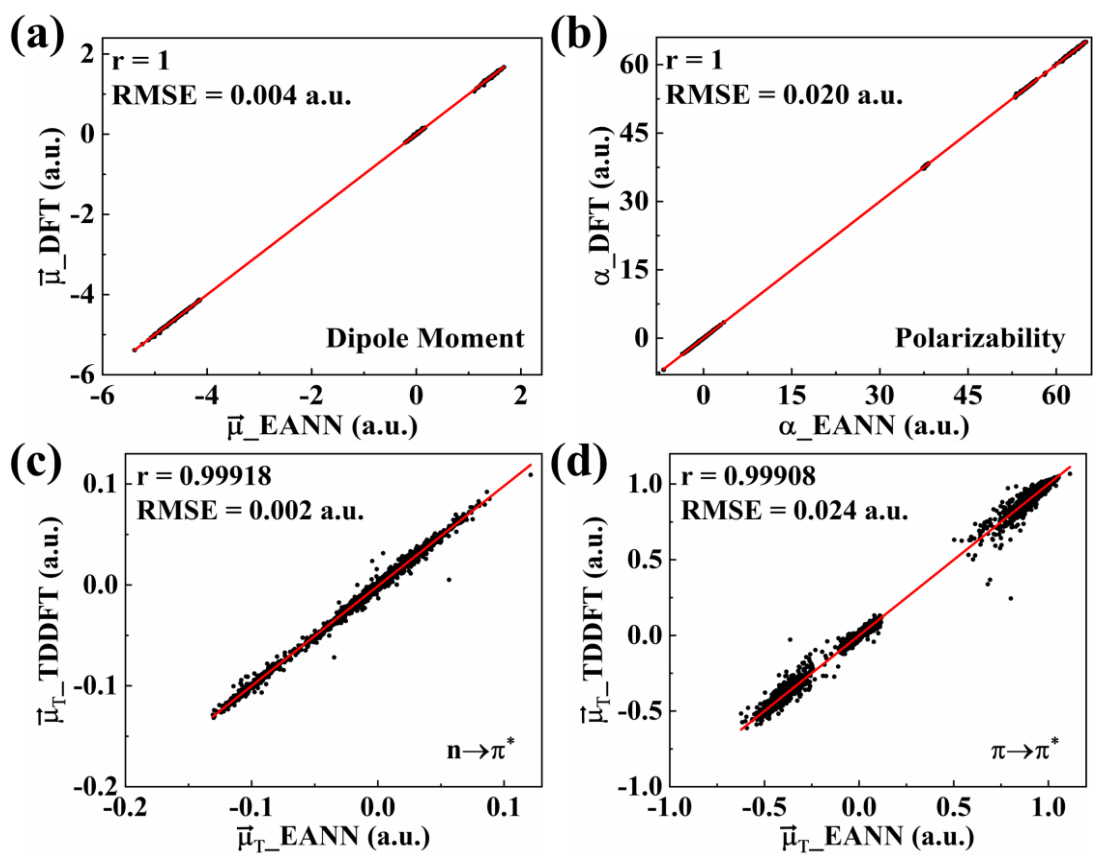

*** Insert Figure 2 ***

Caption: Comparison of ab initio values of (a) dipole moment ($\vec{\mu}$), (b) polarizability ($\boldsymbol{\alpha}$), (c) n→π* and (d) π→π* transition dipole moment ($\vec{\mu}_T$) with the EANN predictions.

Credit: Yaolong Zhang and Bin Jiang

On the other hand, our EANN model is about five orders of magnitude faster than the corresponding DFT and TDDFT calculations. A single DFT calculation of permanent dipole moment and polarizability of NMA takes on average ~1100 s per CPU core (Intel Xeon 6132 2.6 GHz), compared to 9.0 ms using the EANN model with roughly ~$1.2 \times 10^5$ speedup. For transition dipole moments of both n-π* and π-π*, similarly, the EANN and TDDFT computational costs are 3.4 ms versus ~360 s, respectively. These results demonstrate the high accuracy and efficiency of our EANN model for tensorial properties.

[H1 Conclusions and outlook]

This chapter reviews some recent advances on learning dipole moments and related tensorial properties, with special focus on how to reconstruct the physical form of each property to achieve symmetry-preserving and rotational equivariant ML models. In such ways, the modern ML models of learning these tensorial properties can be as accurate and efficient as their scalar counterparts of learning potential energies. There is also rapid progress on the development of unified rotationally-equivariant message-passing neural networks, in which the scalar and tensorial properties can be learned in the same way and different outputs are adapted to corresponding targets[57-59]. This type of model is likely to lead to an increased accuracy in the prediction of tensorial properties (yet also an increase of the computational cost). On the other hand, Ceriotti *et al.* have recently pointed out that it is essential to move beyond the local atomic charge model and include nonlocal effects when describing dipole moments in large molecules whose dipole is almost entirely generated by charge separation[77]. It is worthwhile to note that learning (transition) dipole moments bare much similarity as learning other vectorial properties involving excited states, such as non-adiabatic coupling vectors[51] and spin-orbit coupling vectors[78] (see next Chapter *Learning excited-state properties*). It is also straightforward to apply the model discussed to learn magnetic dipole moments which are useful for predictions of electronic circular dichroism spectra[79]. Some similar ideas and treatments are also invoked in learning electronic friction tensors of adsorbed species on metal surfaces, which have even more complex permutational and rotational equivariance properties[80-82]. Altogether, we look forward to seeing more advances in machine learning of dipole moments and relevant tensorial properties in the near future.

[H1 Acknowledgements]

We acknowledge the continuous support from National Natural Science Foundation of China (22033007), CAS Project for Young Scientists in Basic Research (YSBR-005), Anhui Initiative in Quantum Information Technologies (AHY090200), and K. C. Wong Education (GJTD-2020-15). B. J. is also grateful to the support from National Key R&D Program of China (2017YFA0303500) and The Fundamental Research Funds for the Central Universities (WK2060000017).